\documentclass[prb,twocolumn,showpacs,preprintnumbers,amsmath,amssymb]{revtex4}
\usepackage{graphicx}
\begin{document}
\title{
Dephasing in Quantum Dots: Quadratic Coupling to Acoustic Phonons}
\author{
E. A. Muljarov$^{a,b}$}\email{muljarov@gpi.ru}
\author{R. Zimmermann$^a$ }
\affiliation{ (a) Institut f\"ur Physik der Humboldt-Universit\"at
zu Berlin, Demonstrable 15, D-12489
Berlin, Germany \\
(b) General Physics Institute, Russian Academy of Sciences,
Vavilova 38, Moscow 119991, Russia }
\begin{abstract}

A microscopic theory of optical transitions in quantum dots with
carrier-phonon interaction is developed. Virtual transitions into
higher confined states with acoustic phonon assistance add a
quadratic phonon coupling to the standard linear one, thus
extending the independent Boson model. Summing infinitely many
diagrams in the cumulant, a numerically exact solution for the
interband polarization is found. Its full time dependence and the
absorption lineshape of the quantum dot are calculated. It is the
quadratic interaction which gives rise to a temperature-dependent
broadening of the zero-phonon line, being here calculated for the
first time in a consistent scheme.

\end{abstract}
\date\today

\maketitle

The electron-lattice interaction in quantum dots (QDs) plays a
decisive role for understanding the time evolution and dephasing
mechanisms after optical excitation. High-resolution photoluminescence
spectra on single CdZnTe QDs show a relatively narrow zero-phonon line
(ZPL) with Lorentz broadening $\Gamma$ depending on temperature on top
of a broad band.\cite{Besombes01} The latter is a superposition of
acoustic phonon satellites. Complementary and more detailed information
on the polarization dynamics and dephasing has been obtained from four-wave
mixing measurements in an ensemble of InGaAs QDs.\cite{Borri01} The
polarization decays initially very fast during a few picoseconds
(formation of the broad band), followed by a much longer exponential decay
(tens to hundreds of picoseconds) reflecting the ZPL broadening. This
scenario is referred to in the literature as pure dephasing.\cite{Takagahara99}

In theory, the working horse has been the independent Boson model\cite{Mahan}
which allows an exact analytical solution for the linear optical
polarization\cite{Schmitt87,Besombes01,Krummheuer02,Zimmermann02} and for
the photon echo\cite{Vagov02} of QDs. This model starts with a carrier-phonon
Hamiltonian being linear in the phonon displacement operators, and diagonal
in the confined electronic states. The exact solution
employs the cumulant method and satisfactorily describes the broad phonon
band, i.e. the initial decay of the polarization. However, the ZPL shows no
broadening at all (no longtime decay), in clear contradiction to the
experimental findings. In other words, the phonon cloud induced by
the impulsive excitation of the QD relaxes quickly to a finite lattice
distortion (polaron dressing of the electron-hole pair), but stays constant
during the subsequent time evolution.
To simulate the effect of pure dephasing within the independent
Boson model, a polarization decay can be introduced by hand, or a
phonon damping can be added phenomenologically via the Gr\"uneisen
effect.\cite{Zimmermann02}

To improve on this unsatisfying situation, we develop a purely
microscopic approach to the dephasing in QDs which contains a
finite broadening of the ZPL, thus going essentially beyond the
independent Boson model. We take into account the non-diagonal
terms of the carrier-phonon interaction, i.e. phonon-assisted
transitions of electrons and holes to higher QD
levels.\cite{Takagahara99} As the acoustic phonons which couple to
the QD have energies not more than a few meV, these transitions
are of virtual character only, and do not change the carrier
occupation. Using this argument we apply a unitary transformation
which eliminates the off-diagonal part on the price of adding to
the diagonal one a term being quadratic in the phonon
displacement. Being perturbative, this transformation preserves
the virtue of the independent Boson model, since the new
Hamiltonian allows again a separation of the electron and phonon
coordinates.

With optical phonons, such a quadratic coupling in QDs via virtual
sublevel transitions has been derived recently by Uskov {\it et
al.}\cite{Uskov00} However, the optical response function was
calculated only up to the second order terms in the cumulant,
giving an (approximate) expression for the ZPL decay. If any
damping of the LO phonons is absent, a Gauss decay of the
polarization was found. In contrast, earlier theoretical
work\cite{Levenson71,Osadko72,Hsu84} on impurity-phonon
interaction resulted in an exponential decay, $\exp(-\Gamma t)$,
in agreement with the experiment. The quadratic coupling with
acoustic phonons for the ZPL decay was explored recently by
Goupalov {\it et al.}\cite{Goupalov02} and Hizhnyakov {\it et
al}.\cite{Hizhnyakov02} but assumed to be diagonal in phonon
momentum. This is not correct since the QD breaks the
translational symmetry. All these
approaches,\cite{Levenson71,Osadko72,Hsu84,Goupalov02,Hizhnyakov02}
however, make use of a long-time expansion and are not able to
describe the initial rapid decay (the broad band).

In this contribution we find an {\it exact solution} of the
carrier-phonon problem in QDs with quadratic coupling to acoustic
phonons. We calculate for the first time the full time-dependent
optical polarization and the absorption spectrum of a QD,
including both the broad band and the finite broadening of the
ZPL.

After eliminating the inter-sublevel coupling
the transformed Hamiltonian has the following form
\begin{eqnarray}
\hat{H}& = &\sum_{\bf q} \hbar\omega_{\bf q} a^+_{\bf q} a_{\bf q}
 \, + \, (\hbar\omega_{eh}+\hat{V})|1\rangle\langle 1|,
\label{Hamiltonian} \\
\hat{V}& = &\sum_{\bf q} M_{\bf q}(a_{\bf q} + a^+_{-\bf q})
 \nonumber\\
 &&-\sum_{{\bf q q}'} Q_{{\bf q q}'}(a_{\bf q}+ a^+_{-\bf
q})(a_{{\bf q}'} + a^+_{-{\bf q}'}), \label{V}
%%\\
\end{eqnarray}
%%\begin{eqnarray}
\begin{equation}
M_{\bf q}=M^{11}_{{\bf q}e}-M^{11}_{{\bf q}h} \, , \ \ \ Q_{{\bf q
q}'}=\sum_{a=e,h} \sum_{\nu\neq 1}\frac{M^{1\nu}_{{\bf q}a}M^{\nu
1}_{{\bf q}'a}}{E_a^\nu-E_a^1}. \label{Qqq}
\end{equation}
%%\end{eqnarray}
%
The electronic Hilbert space consists of the QD ground state
$|0\rangle$ and the excited state $|1\rangle$ having just one
electron in the lowest conduction-band level ($E_e^1$) and one
hole in the uppermost valence-band level ($E_h^1$). The bare
transition energy is denoted by $\hbar\omega_{eh}=E_g + E_e^1 +
E_h^1$. The matrix element for the deformation potential coupling
with longitudinal acoustic phonons in the QD reads ($a = e, h$)
\begin{eqnarray}     \label{Defpot}
M^{\nu\nu'}_{{\bf q} a}& = &\sqrt{\frac{\hbar\omega_q}{2\rho_M
u^2_s V}} D_a \int d{\bf r} \,\psi_{\nu a}^\ast({\bf
r}) e^{i{\bf qr}} \psi_{\nu'a}({\bf r}) \, ,
\end{eqnarray}
with $\psi_{\nu a}({\bf r})$ and $D_a$ being the confinement wave
function and the deformation potential constant, respectively.
$\rho_M$ is the mass density, $u_s$ the sound velocity, and $V$
the phonon normalization volume. In writing Eq.(\ref{Defpot}) we
have neglected (i) the excitonic Coulomb correlation which is of
minor importance for small QDs\cite{Zimmermann02} but could be
included easily; (ii) any difference between phonon parameters in
the QD and in the barrier material leading to confined modes, thus
dealing with bulk phonons only; (iii) deviations of the acoustic
phonon dispersion from a linear isotropic one, $\omega_{\bf q}=u_s
q$.

The linear response function, i.e. the polarization after a
delta-pulse excitation at $t=0$, is given by the dipole-dipole
correlation function $P(t) = i\,\langle d^\dagger(t) \, d(0)\rangle$
for $t>0$,
\begin{eqnarray}  \label{polarization}
P(t)& = &i\,e^{-i\omega_{eh}t}\left\langle {\cal T}\exp\left[
-\frac{i}{\hbar}\int_0^td\tau\hat{V}(\tau) \right]\right\rangle \\
& = &i\,e^{-i\omega_{eh}t}
\exp\left[\sum_{n=1}^\infty\left(\frac{-i}{\hbar}\right)^n\frac{1}{n!}
 \int_0^t dt_1 \dots dt_n \right.
 \nonumber\\
&&\left.\times\Bigl\langle {\cal T} \hat{V}(t_1) \dots
 \hat{V}(t_n)\Bigr\rangle_{\rm conn} \right] \, , \nonumber
\end{eqnarray}
with $\cal T$ being the time ordering, and using the interaction
representation of operators. The expectation value for the
electronic system is already performed, leaving only the average
over the phonon operators to be done. In the second line of
Eq.~(\ref{polarization}) we have used the cumulant
expansion\cite{Mahan} which has only connected diagrams in the
exponent. Applying Wick's theorem, the thermal phonon average
forces the displacement operators to appear only pairwise in
expectation values, which defines the (momentum diagonal) phonon
propagator
\begin{eqnarray}
D_{\bf q}(t) & = & (-i/\hbar)
\left\langle {\cal T}\left[a_{\bf
q}(t) + a^\dagger_{\bf q}(t)\right]
\left[a_{\bf q}(0) + a^\dagger_{\bf q}(0)\right]\right\rangle \nonumber \\
& = & (-i/\hbar)\left[(N_{\bf q}+1) e^{-i\,\omega_{\bf q}|t|}+
N_{\bf q} e^{i\,\omega_{\bf q}|t|}\right]
\end{eqnarray}
as building block. The phonon occupation enters via the Bose
function, $ N_{\bf q}=1/[\exp(\hbar\omega_{\bf q}/k_B T)-1]$. In
the cumulant technique, connected diagrams are those which do not
factorize. Thus, the propagators have to appear in different time
integrations, forming a chain-like structure. Concentrating first
on the linear interaction, there is only one diagram which is
connected [$n = 2$ in Eq.~(\ref{polarization})]. In all higher
orders, the phonon propagators have their time integrations
exclusively, and the expressions do factorize. Things are quite
different for the quadratic interaction since now two displacement
operators share the same time argument (but have different
momentum). To form a connected diagram, they have to be
distributed on two different propagators, ending up with a closed
loop. Finally, a combination of linear and quadratic interactions
adds another connected structure, which is a chain with two open
ends. The linear coupling can appear only twice, just at the ends.

The calculation method for the diagram series developed in this
work can be applied to any well-behaved function $Q_{{\bf qq}'}$.
Nevertheless, a great numerical simplification can be achieved if
a simple factorization of the dependence on ${\bf q}$ and ${\bf
q'}$ holds. Physically, this case may appear if one excited level
having the smallest energy distance dominates the sum in
Eq.~(\ref{Qqq}). Concentrating at the moment on the next hole
level $E^2_h$, we have with $\Delta_h=E^2_h-E^1_h$ the product
form $Q_{{\bf qq}'}= M^{12}_{{\bf q}h} M^{21}_{{\bf
q'}h}/\Delta_h$. Then, summing over momentum, only three different
propagators appear,
\begin{eqnarray} \label{new-D}
&&D_L(t)=1/\hbar \sum_{\bf q}|M_{\bf q}|^2 D_{\bf q}(t) \, ,
\nonumber\\
&&D_Q(t)=-2/\Delta_h \sum_{\bf q}|M^{12}_{{\bf q} h}|^2 D_{\bf q}(t) \, ,\\
&&D_M(t)=\sqrt{2/(\hbar\Delta_h)}\sum_{\bf q} M_{\bf q}^\ast
  M^{12}_{{\bf q} h}D_{\bf q}(t) \, . \nonumber
\end{eqnarray}
As a result we obtain the cumulant expansion
\begin{equation}
P(t)=i\,\exp[-i\omega_{eh}t+K_L(t)+K_Q(t)+K_M(t)]
\end{equation}
with
\begin{eqnarray} \label{LQM}
K_L(t)&=&-\frac{i}{2} \int_0^t dt_1 dt_2 D_L(t_1-t_2) \, , \nonumber\\
K_Q(t)&=&\frac{1}{2}\sum_{n=1}^\infty \frac{1}{n}\int_0^t
dt_1\dots dt_n \\
&&\times D_Q(t_1-t_2)\dots D_Q(t_n-t_1) \, , \nonumber\\
K_M(t)&=&\frac{i}{2}\sum_{n=1}^{\infty} \int_0^t dt_0\dots
dt_{n+1} D_M(t_0-t_1) \nonumber \\
&\times& D_Q(t_1-t_2)\dots
D_Q(t_{n-1}-t_n) D_M(t_{n}-t_{n+1}) \, . \nonumber
\end{eqnarray}
The first term in Eq.\,(\ref{LQM}), $K_L$, is all what remains from the
linear coupling, since only one diagram survives as discussed above.
This reproduces exactly the result within the independent Boson model.
The quadratic interaction brings in a purely quadratic term, $K_Q$, and a
mixed one, $K_M$, which both contain an infinite sum over diagrams. Note
that the factorial in Eq.\,(\ref{polarization}) is reduced to $1/n$ and 1,
respectively, due to a proper counting of equivalent diagrams.

For realistic coupling strengths, a calculation of the sum over $n$ term
by term is hopeless. We follow another path which allows us to sum up
the infinite sum numerically exact: With the quadratic propagator $D_Q(t)$
as kernel, we look for the eigenvalue problem of the Fredholm integral
equation
\begin{equation}
\int_0^t dt_2 \, D_Q(t_1-t_2) \, u_j(t_2) = \Lambda_j \, u_j(t_1) \, .
\end{equation}
The kernel is complex and symmetric, but not Hermitian. Therefore,
the eigenvalues $\Lambda_j (j=1,2,\cdots)$ are complex, but the
eigenfunctions still form an orthonormal system. Note that both
eigenvalues and eigenfunctions depend parametrically on the actual
time $t$ which defines the integration limits. Expanding the
multiple integrals into the orthonormal system, an $n$-fold
convolution of $D_Q(t)$ converts into a power $\Lambda_j^n$. The
subsequent summation over $n$ results in a logarithm for the
quadratic term, and leads to a geometrical series for the mixed
one,
\begin{eqnarray}
K_Q(t) & = & -\frac{1}{2} \sum_j \ln (1-\Lambda_j) \, ,
\label{cumQ}\\
K_M(t) & = & \frac{i}{2} \sum_j \frac{D^2_{Mj}}{1-\Lambda_j},
\label{cumM}\\
\mbox{where} && D_{Mj} = \int_0^t dt_1 dt_2 D_M(t_1-t_2) \, u_j(t_2)
 \, . \nonumber
\end{eqnarray}
Technically, we have discretized the time integrations on a fine grid, thus
converting the Fredholm problem to a standard matrix eigenvalue search.

Up to now we have considered virtual transitions into one excited
hole level only. It can be shown, however, that the factorization
of $Q_{\bf q q'}$ can be still employed for several excited levels
if they refer to different spatial symmetries. E.g., if the
quantum dot has mirror symmetry, several excited odd states of
different orientation have vanishing cross terms in the quadratic
cumulant $D_Q(t)$, and contribute additively to $K_Q(t)$.
According to the same argument, $D_M(t) \equiv 0$ and consequently
$K_M(t) \equiv 0$. In the actual calculation we assume a spherical
quantum dot with parabolic confinement potential.  The wave
functions of the lowest states, $\psi_{1a}({\bf r})$, are
isotropic Gauss orbitals with variance $l_a$. Taking artificially
$l_e = l_h=l$ allows us to take into account the next three
(degenerate) excited states for both, electron and hole
simultaneously, without spoiling the factorization scheme. The
only changes are the replacement $D_v^2/\Delta_h\to
D_c^2/\Delta_e+D_v^2/\Delta_h$ in the definition of $D_Q(t)$, and
an additional factor of three in $K_Q(t)$, Eq.~(\ref{cumQ}).

In the calculations, we have taken InAs parameters\cite{Landolt}
$\rho_M=5.67$\,g/cm$^3$, $u_s=4.6\!\cdot\! 10^3$\,m/s,
$D_{c}=-13.6$\,eV, $D_c-D_{v}=-6.5$\,eV. The confinement variance
$l$ is the only parameter characterizing the QD. The matrix
element Eq.(\ref{Defpot}) behaves like $(M^{11}_{{\bf q}a})^2
\propto q \exp[-(q l)^2/2]$, which gives for the different
propagators a temporal decay on the scale $l/u_s$. Consequently,
$\hbar u_s/l$ sets the scale for the phonon energies involved, and
gives roughly the spectral width of the phonon satellites (broad
band). For the QDs studied in Ref.~\onlinecite{Borri01}, we have
extracted a value of $l \approx 3.3$\,nm. In a spherical QD with
parabolic confinement, the distance to the next sublevel is given
by $\Delta_a = \hbar^2/(m_a l^2)$. To come closer to the
experimental situation of a more pancake shaped quantum dot, we
have instead taken the experimental observation of $\Delta_e +
\Delta_h = 65\,$meV and distributed between electron and hole
according to the InAs mass ratio of 10:1.

\begin{figure}[t]
\includegraphics*[width=8cm]{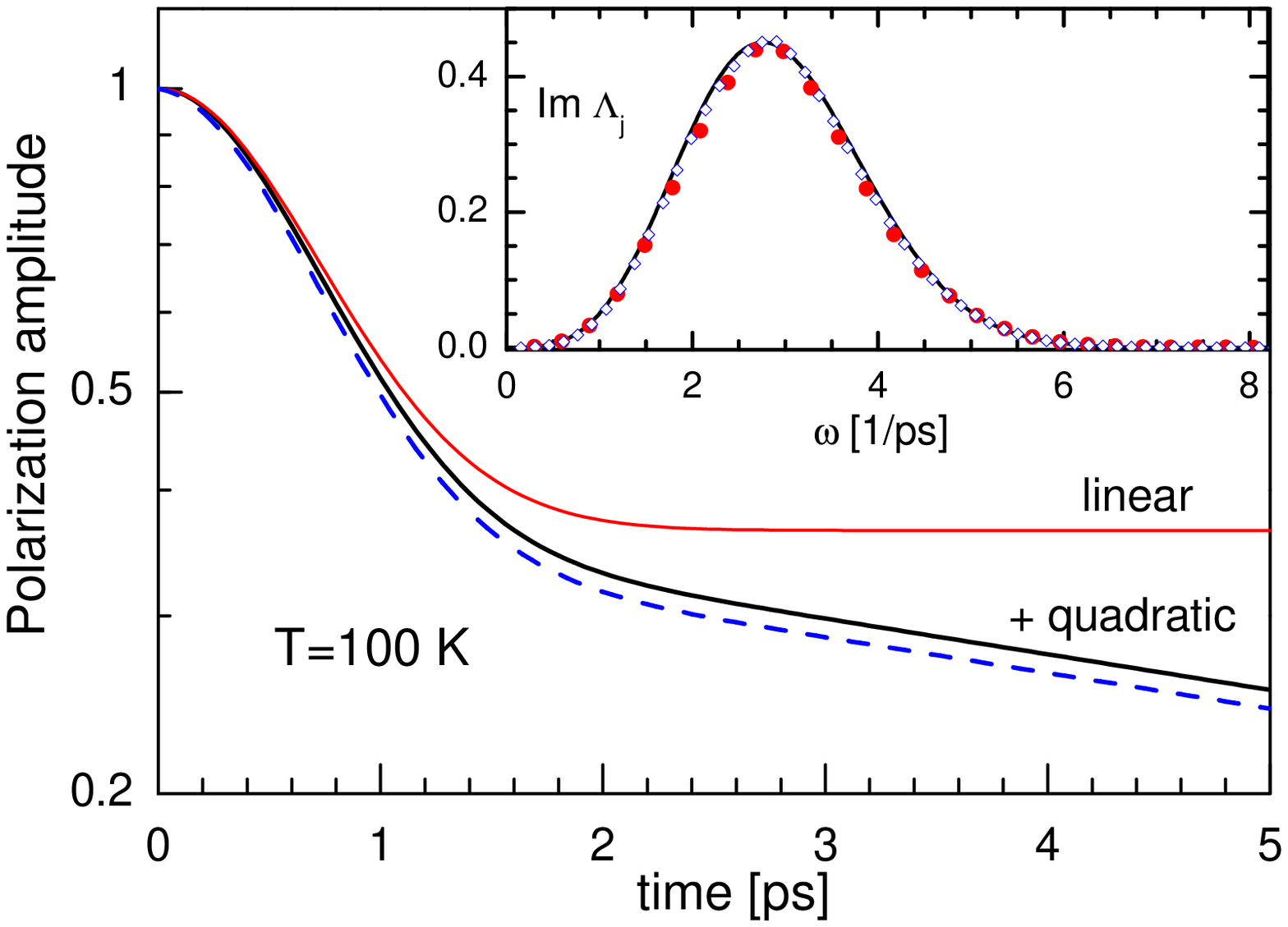}%
\caption{\label{fig:PolarLa} Polarization amplitude calculated for
an InAs QD with $l=3.3$\,nm at $T=100$\,K with linear and
quadratic coupling to acoustic phonons. For the dashed curve, a
finite decay $1/\Gamma = 12.3$\,ps has been added by hand to the
linear result. The inset shows the (dominant) imaginary part of
the eigenvalues $\Lambda_j$ at $\omega_j$ for $t = 10$\,ps
(circles) and 20 \,ps (diamonds), superimposed to the Fourier
transform $\tilde{D}_Q(\omega)$ (solid curve).}
\end{figure}

The calculated polarization amplitude in Fig.~\ref{fig:PolarLa}
undergoes a quick initial decay on the scale $l/u_s \approx
1$\,ps, which reflects the formation of the polaron cloud around
the e-h-pair. This state, however, is not completely stable: Due
to the quadratic coupling, there are virtual scattering events
into higher states which distort the polarization, leading to an
exponential decay (pure dephasing). Note that the quadratic
coupling does not change much the initial decay, as seen from the
dashed curve in Fig.~\ref{fig:PolarLa} where a decay is added by
hand to the linear result.

For times much larger than the initial decay ($\propto l/u_s$) a
more detailed analysis is possible. The linear term alone gives
$K_L(t) \rightarrow -S_L + i\Omega_L t$, which is a reduction of
the polarization amplitude $\exp(-S)$ ($S$ is the Huang-Rhys
factor) and a purely imaginary contribution linear in time
(polaron shift $\Omega$). The quadratic term can be analyzed in
terms of the eigenvalues $\Lambda_j$. For $t$ larger than the
temporal width of $D_Q(t_1)$, the eigenfunctions are found to
oscillate in the middle of the interval as $u_j(t_1) \propto
\exp(i\omega_j t_1)$, while $\Lambda_j = \tilde{D}_Q(\omega_j)$
with the Fourier transform of $D_Q(t_1)$. The exact position of
the eigenvalues (or real frequencies $\omega_j$) is determined by
the boundary conditions at $t_1 = 0$ and $t_1 = t$. Numerically,
we found very good agreement with
\begin{equation}  \label{Lam-j}
\Lambda_j(t) \approx \tilde{D}_Q\left(\omega_j = \frac{\pi
j}{t+2\tau}\right) \, .
\end{equation}
The temporal shift $\tau$ is practically independent on $t$ and takes care of
the `border correction' from the finite integration interval.

In the inset of Fig.~\ref{fig:PolarLa} we plot the eigenvalues at
two different times, adjusting $\tau = 0.27\,$ps for optimal
agreement with $\tilde{D}_Q(\omega)$ (solid curve). Notice that
for $t=20$\,ps the eigenvalues are two times denser than for
$t=10$\,ps, as dictated by Eq.\,(\ref{Lam-j}). Combining
Eqs.\,(\ref{cumQ},\ref{Lam-j}) we \mbox{arrive at}
\begin{eqnarray}  \label{K-asym}
K_Q(t) & = & -(t+2\tau)\int_0^\infty\frac{d\omega}{2\pi}\,
\ln\left[1-\tilde{D}_Q(\omega)\right] \nonumber \\
& = & -S_Q - \Gamma t + i\Omega_Q t
\end{eqnarray}
which has now a {\it real} part in the linear time dependence.
This gives the polarization dephasing or ZPL width in the frequency domain.
These analytical expressions for dephasing rate
$\Gamma$ and frequency shift $\Omega_Q$ have been derived earlier
in different models of the quadratic electron-phonon
coupling.\cite{Levenson71,Osadko72,Hsu84,Goupalov02,Hizhnyakov02}
New is the inclusion of the time-independent correction $\propto \tau$
which makes the Huang-Rhys factor $S = S_L + S_Q$ a complex quantity,
thus adding to the ZPL Lorentzian a (small) dispersive contribution.

\begin{figure}[t]
\includegraphics[width=8cm]{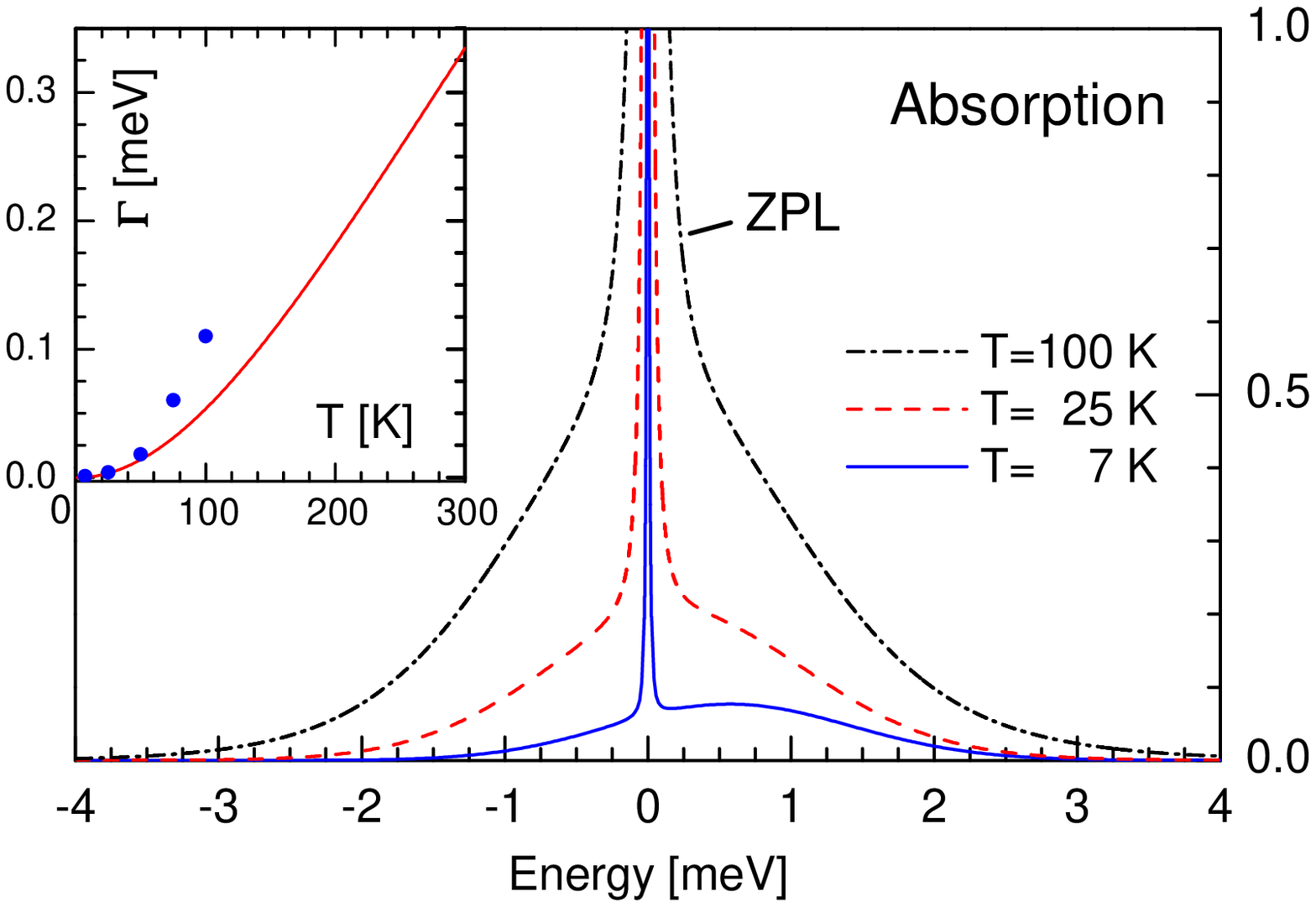}%
\caption{\label{fig:SpectGa}Absorption spectra of an InAs QD with
carrier confinement length of $l=3.3$\,nm at different
temperatures. The ZPL transition energy is taken as zero of
energy. Inset: Calculated broadening of the ZPL compared with the
experimental results by Borri et.al.\cite{Borri01} (circles).}
\end{figure}

From the complex polarization follows the absorption spectrum
easily, $\alpha(\omega)={\rm Im} \int_0^\infty d t\,P(t)
\exp(i\omega t)$ (Fig.~\ref{fig:SpectGa}). At low temperatures,
the broad band is distinctly asymmetric and gets more symmetric as
$T$ grows. This is not much different from the results of linear
coupling. However, in accordance with the decay seen in
Fig.~\ref{fig:PolarLa}, the ZPL acquires a Lorentz broadening with
width $\Gamma(T)$, which is exclusively due to the quadratic
coupling. The inset shows the calculated temperature dependence of
the dephasing rate $\Gamma(T)$ (solid curve). Starting with
$\Gamma(0) = 0$, the full curve has nearly parabolic shape before
tending to a more linear temperature dependence. The comparison
with the experimental data (circles) by Borri {\it et
al.}\cite{Borri01} shows that a quantitative agreement has not
been reached. We attribute this to the restriction of virtual
transitions into the next degenerate dot states. An approximate
estimate for the full sum over levels gives indeed a substantial
increase. Then, however, states in the wetting layer will be
important as well.

In conclusion, we have developed a microscopic theory of dephasing
in quantum dots accounting for quadratic coupling between carriers
and acoustic phonons. Introducing a new numerical method allowed
us to perform the infinite sum of diagrams in the cumulant. This
exact solution gives the full time dependence of the optical
polarization. The absorption spectrum includes the broad band and
the broadening of the zero-phonon line on the same footing. Our
calculated results for the ZPL decay in InGaAs quantum dots show
the same trend as the experiment.\cite{Borri01}\\

%\begin{acknowledgments}
Financial support by DFG (Sfb 296), DAAD (NATO fellowship 325-A/03/06772),
and the Russian Foundation for Basic Research is gratefully acknowledged.
%\end{acknowledgments}

\end{document}